\documentclass[twocolumn,prl,aps,showpacs,superscriptaddress]{revtex4}
\usepackage{epsfig}
\usepackage{bm,hyperref}
\usepackage{amssymb}

\def\ha{\hat{a}}
\def\had{\hat{a}^{\dagger}}
\def\hb{\hat{b}}
\def\hbd{\hat{b}^{\dagger}}

\def\hh{\hat{H}}
\def\r{\bm R}

\def\sem{\hbar\rightarrow 0}
\def\adia{\alpha\rightarrow 0}
\def\semi2{N\rightarrow \infty}

\def\be{\begin{equation}}
\def\ee{\end{equation}}
\def\ba{\begin{eqnarray}}
\def\ea{\end{eqnarray}}

\begin{document}
\title{Commutability between Semiclassical Limit and Adiabatic Limit
}
\author{Biao Wu}
\affiliation{Institute of Physics, Chinese Academy of Sciences, P.O. Box 603, 
Beijing 100080, China}
\author{Jie Liu}
\affiliation{Institute of Applied Physics and Computational Mathematics, 
P.O. Box 8009, Beijing 100088, China}
\date{July 30th, 2005}

\begin{abstract}
We study the adiabatic limit and the semiclassical limit with a second-quantized
two-mode model, which describes a many-boson interacting system.
When its mean-field interaction is small, these two limits are commutable.
However, when the interaction is strong and over a critical value, the 
two limits become incommutable. This change of commutability is associated
with a topological change in the structure of the energy bands.
These results reveal that nonlinear mean-field theories, such as Gross-Pitaevskii 
equations for Bose-Einstein condensates, can be invalid in the adiabatic limit.
\end{abstract}

\maketitle
The experimental creation of Bose-Einstein condensates (BECs) with 
dilute alkali atomic gases has generated great excitement and literally 
created a new subfield in physics\cite{bec_rmp,leggett}. One of the main 
reasons is that it has made it possible to test experimentally some fundamental 
and important physics that could only be discussed theoretically before. 
For instance, Tonks-Girardeau gas and quantum phase transition between 
superfluid and Mott insulator has been studied by theorists since 1960s;
they were observed experimentally only recently with BECs\cite{mott,tonks}. 
There are now even discussions on how to use BECs to study 
black holes\cite{blackhole} and  superstrings\cite{string}.

In this Letter we discuss a fundamental concept in quantum mechanics,
the commutability of the semiclassical limit and the adiabatic 
limit, with a second-quantized two-mode model. We suggest a possible 
experimental testing of this concept with BECs. This concept is 
due to M.V. Berry\cite{berry}. In brief, consider a quantum system whose 
Hamiltonian is time-dependent,
\be
\label{eq:sch}
i\hbar\frac{\partial}{\partial t}|\psi\rangle=H(\r(\alpha t))|\psi\rangle\,.
\ee
One can eliminate $\alpha$ from the above Schr\"odinger 
equation with a scaled time $\tau=\alpha t$ and an effective Plank 
constant $\widetilde{\hbar}=\alpha\hbar$. 
With this scaling argument, Hwang and Pechukas claimed that the 
semiclassical limit $\sem$ and the adiabatic limit $\adia$ are 
equivalent\cite{hwang}.

This point was refuted by Berry\cite{berry}, who pointed out that these two 
limits are not equivalent because the Hamiltonian $H$ may depend implicitly 
on $\hbar$. Moreover, he showed that these two limits are incommutable in 
a simple double-well model: the Landau-Zener (LZ) 
tunneling rate in this model is zero if the adiabatic limit 
$\adia$ is taken first; it becomes one when the semiclassical limit $\sem$ 
is taken first\cite{berry}. Since it is impossible to change $\hbar$ 
experimentally, this concept has remained a game of theorists.

We revisit the commutability between the semiclassical limit and  the
adiabatic limit with a second-quantized two-mode tunneling model. This model 
can be used to describe a BEC system where only two quantum states are important, 
such as in a double-well potential or with two internal quantum 
states\cite{leggett,smerzi}. In this model,  the semiclassical limit
is $\semi2$ with $N$ being the number of bosons. In this large $N$ limit,
the second-quantized model becomes a two-level mean-field model. We show
that one can recover the second-quantized model by quantizing this
mean-field model with the Sommerfeld rule. As $N$ can be changed 
in experiments, the semiclassical limit becomes experimentally accessible.
 
More interestingly, the commutability between the two limits, $\semi2$ 
and $\adia$, in this second-quantized model depends on its mean-field 
interaction strength $c$. If $c$ is small, the two limits are commutable; 
when $c$ is over a critical value, the two limits become incommutable. Such 
a dependence on $c$ is found to be related to a topological change 
in the structure of the energy bands. These results indicate that 
nonlinear mean-field theories, such as Gross-Pitaevskii equations for BECs, 
can be invalid in the adiabatic limit when the mean-field interaction 
is strong.  Finally, we discuss how this commutability 
can be tested in a BEC experiment.

The second-quantized  two-mode model is 
\be
\label{eq:qham}
\hh={\gamma\over 2}(\had\ha-\hbd\hb)+
{v\over 2}(\had\hb+\ha\hbd)-{\lambda\over 4}
(\had\ha-\hbd\hb)^2\,,
\ee
where generators and annihilators $\had,\ha$ and $\hbd,\hb$ are for 
two different quantum states. 
In the Hamiltonian $\hh$, $\gamma$ is the energy offset between the two 
quantum states and changes with time as $\gamma=\alpha t$. The parameter $v$ 
measures the coupling between the two states while $\lambda>0$ is the 
interacting strength between bosons.
The minus sign before $\lambda$ indicates that the interaction is attractive. 
In this system the total number of bosons $N$ is conserved.

For this second-quantized model, its semiclassical limit is $\semi2$. 
In such a limit, the system's dynamics is given by the following
nonlinear two-level model,
\be
\label{eq:mham}
i\frac{d}{dt}\pmatrix{a\cr b}=\Big\{[\frac{\gamma}{2}-\frac{c}{2}(|a|^2-|b|^2)]\sigma_z
+\frac{v}{2}\sigma_x\Big\}\pmatrix{a\cr b}\,.
\ee
where $c=N\lambda$ and $|a|^2+|b|^2=1$. This model is often called a mean-field 
model. Technically to obtain the mean-field model, one focuses on 
the Gross-Pitaevskii states\cite{leggett}
$
|\Psi_{gp}\rangle=\frac{1}{\sqrt{N!}}(a\had+b\hbd)^N|{\rm vac}\rangle\,.
$
By computing the expectation value 
$\langle\hh\rangle=\langle \Psi_{gp}|\hh|\Psi_{gp}\rangle$, one obtains
the mean-field Hamiltonian $H_{\rm mf}=\langle\hh\rangle/N$ 
(up to a trivial constant) in the limit of $\semi2$. The Hamiltonian $H_{\rm mf}$
leads to the dynamics in Eq.(\ref{eq:mham}).
For a rigorous account of large $N$ limit as a semiclassical limit
in models such as Eq.(\ref{eq:qham}), we refer readers to Ref.\cite{zhang}.

We emphasize  that the semiclassical limit $\semi2$ is taken with 
the mean-field interaction strength $c=N\lambda$ kept constant. 
Physically, this is to ensure that the series of systems with different 
$N$'s have about the same physics. If $\lambda$ were kept constant 
instead of $c$, the last term in Eq.(\ref{eq:qham}) 
would become too dominating at the large $N$ limit, completely changing the
physics of the system. When the model (\ref{eq:qham})
is used to describe a BEC in a double-well potential, the limit $\semi2$
at a constant $c$ is equivalent to having a larger trap 
for more atoms in the BEC, or to tuning $\lambda$ smaller with the
Feshbach resonance technique\cite{feshbach}.

We are interested in how the second-quantized model Eq.(\ref{eq:qham})
behaves in the two limits, $\semi2$ and $\adia$, in particular,
whether the model's behavior depends on which limit
is taken first. For this purpose, we follow Berry's methodology\cite{berry}
to focus on the tunneling behavior of the quantized model.

\begin{figure}[!htb]
\includegraphics[width=7.0cm]{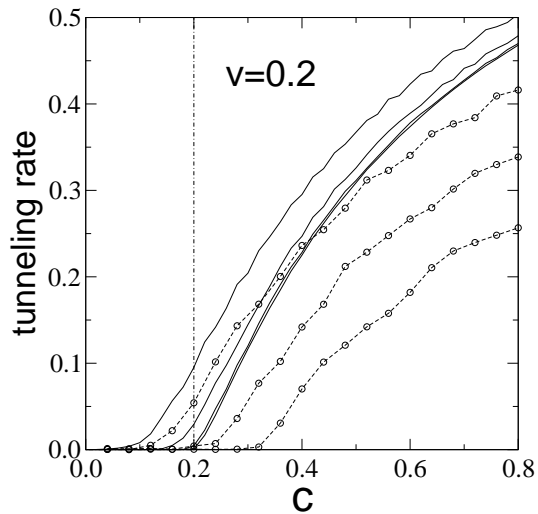}
\caption{Tunneling rate as a function of the mean-field interaction strength $c$.
The solid lines are obtained with the mean-field model (\ref{eq:mham}) for
$\alpha=0.005,0.001,0.0001,0.00001$(from top to bottom);
the circle-dashed lines are with the quantized model (\ref{eq:qham})
for $\alpha=0.005,0.001,0.0001$(from top to bottom). $N=8$ and $v=0.2$
are used.}
\label{fig:trate}
\end{figure}

In Fig.\ref{fig:trate} the tunneling rates are plotted as a function of the 
mean-field interaction strength $c$. Two sets of tunneling rates are calculated:
one with the quantized model (\ref{eq:qham}) for a fixed number of bosons;
the other with the mean-field model (\ref{eq:mham}). 
In computing the tunneling rate, we have assumed that the system is completely 
in state $a$ at $t\rightarrow -\infty$; the tunneling rate is the probability of 
remaining in state $a$ at $t\rightarrow \infty$, the end of dynamical evolutions. 
At a fixed number of bosons, the dynamics of the quantized model (\ref{eq:qham})
can be found by expanding a quantum state in terms of Fock states $|N_a,N_b\rangle$,
where $N_a$ and $N_b$ are numbers of particles in quantum states $a$ and $b$,
respectively. In terms of $|N_a,N_b\rangle$, the Hamiltonian becomes a
$(N+1)\times(N+1)$ matrix.

Upon careful examination of the data in Fig.\ref{fig:trate},
one notices that $c=v$ is a critical value. When $c<v$, the tunneling rate 
goes to zero in the adiabatic limit $\adia$ for both the mean-field model
and the quantized model. However, when $c>v$, the tunneling rate from 
the mean-field model is always non-zero while the tunneling rate can be zero 
for the quantized model. Since the mean-field model is the semiclassical 
limit of the quantized model, the mean-field result can be regarded as 
the result from the quantized model with the limit $\semi2$ having been taken.
Therefore, the results in Fig.\ref{fig:trate} show that the tunneling
behavior in the quantized model (\ref{eq:qham}) depends strongly
on the order of the  limits taken while this dependence itself
relies on the value of the mean-field interaction strength $c$.

\begin{figure}[!htb]
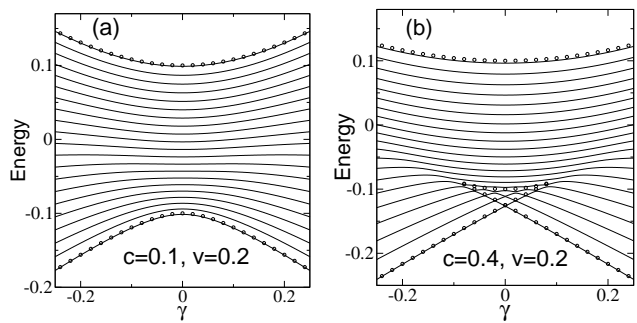

\includegraphics[width=4cm]{qnlz1}\hspace{0.2cm}
\includegraphics[width=4cm]{qnlz3}
\caption{Energy levels from the second-quantized model $\hh$ ($N=20$) and the 
mean-field model $H_{\rm mf}$. The solid lines are quantized energy levels; 
the open circles are mean-field energy levels. Note that for comparison 
with the mean-field theory, the quantized energy levels from $\hh$ have been 
divided by $N$.}
\label{fig:qnlz}
\end{figure}

To understand the above results, we first examine the energy levels of 
the second-quantized model (\ref{eq:qham}) as functions of $\gamma$, the slowly 
changing system parameter. These energy levels can be found by directly
diagonalizing the Hamiltonian $\hh$ and they are plotted in Fig.\ref{fig:qnlz}.
There is a drastic change in the structure of energy 
levels as the mean-field interaction $c$ changes:  a net of anti-crossings appearers 
in the lower part of the quantized energy levels when $c>v$. As known before\cite{nlz}, 
when $c>v$ there is a loop structure emerging in the energy band of the 
mean-field model (\ref{eq:mham}). When the mean-field energy 
levels (circles) are also plotted in Fig.\ref{fig:qnlz}, we find that the quantized 
energy levels are bounded by the mean-field energies. In particular, the 
mean-field energy levels envelop the net of anti-crossings in the quantized 
energy levels. Such a correspondence was first noticed in Ref.\cite{smerzi}.

\begin{figure}[!htb]
\includegraphics[width=7.5cm]{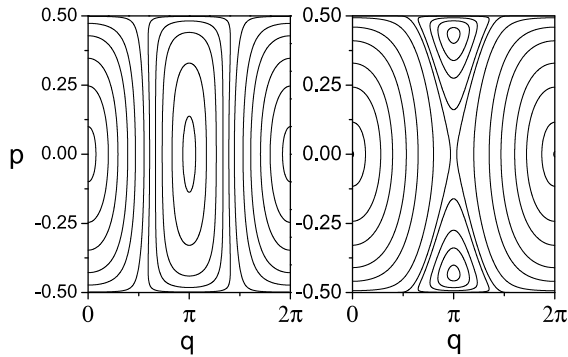}
\caption{Energy contours of the mean-field model $H_{\rm mf}$.
Left: $c=0.1$, $v=0.2$, $\gamma=0.0$; right:$c=0.4$, $v=0.2$, $\gamma=0.0$.}
\label{fig:england}
\end{figure}

The structure change in the energy bands is associated with a change in 
the phase space of the mean-field model (\ref{eq:mham}) as shown in
Fig.\ref{fig:england}. In plotting this figure, we notice that 
the mean-field model, in fact, has only two independent variables and
its Hamiltonian can be reduced to
\be
\label{eq:hampq}
H_{\rm mf}=\gamma p+\frac{v}{2}\sqrt{1-4p^2}\cos q-cp^2\,,
\ee
where $ p=(|a|^2-|b|^2)/2$ and $q=\theta_b-\theta_a$
with $\theta_{a,b}$ being the phases of $a$ and $b$. It is clear from 
Fig.\ref{fig:england}, when $c<v$, there is one minimum 
and one maximum; when $c>v$, we see two local minima, one maximum and 
one saddle point. Since these extremum points correspond to the eigenstates 
in the mean-field energy bands in Fig.\ref{fig:qnlz}\cite{liu},
the structure change in the phase space is apparently connected
with the structure change in the energy levels.

\begin{figure}[!ht]
\includegraphics[width=7.5cm]{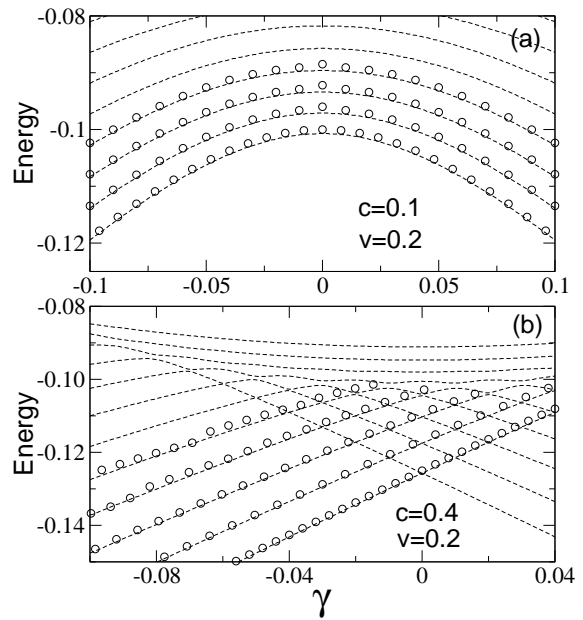}
\caption{Comparison between the energy levels of the second-quantized 
model (dashed line) with $N=40$ and the Sommerfeld energy levels (open circles). 
(a) $c=0.1$, $v=0.2$; (b) $c=0.4$, $v=0.2$. For clarity, we have plotted 
only a portion of the energy levels.}
\label{fig:quant}
\end{figure}

This connection can be further  explored by re-quantizing the mean-field model 
$H_{\rm mf}$ with the Sommerfeld theory, which says that the quantum 
motions are the periodic motions in the classical phase space that satisfy
\be
\label{eq:sommer}
\frac{1}{2\pi}\oint pdq=n\hbar/N\,,\hspace{1cm}n=0,1,2,\cdots
\ee
The division by $N$ comes from the fact that the mean-field Hamiltonian
is an average for one particle, $H_{\rm mf}=\langle\hh\rangle/N$.
One can view $\hbar_{eff}=\hbar/N$ as the effective Plank constant 
for $H_{\rm mf}$. In our calculations, the natural unit $\hbar=1$ is used.
For convenience, we shall call the energy levels obtained with Eq.(\ref{eq:sommer})
the Sommerfeld energy levels. They are 
shown and compared to the quantized energy levels of $\hh$ 
in Fig.\ref{fig:quant}.

\begin{figure}[!htb]
\includegraphics[width=6.5cm]{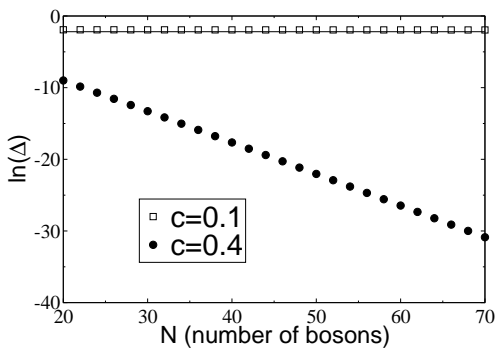}
\caption{Energy gap between the lowest two eigen-energies
in the second-quantized model at $\gamma=0$. The squares
are for $c=0.1$ and the dots for $c=0.4$, with $v=0.2$
for both. The solid line is an approximation result 
$\Delta=v\sqrt{1-c^2/v^2}$ for $c<v$.}
\label{fig:gap}
\end{figure}

When $c<v$, the mean-field Hamiltonian has exactly one maximum ($q=0$) and 
one minimum ($q=\pi$). The Sommerfeld quantization around the maximum produces
energy levels lower than the maximum energy while the quantization
around the minimum generates energy levels higher than the minimum.
This explains why the mean-field energy levels bound the quantized energy
levels in Fig.\ref{fig:qnlz}. We also see that the energy gap arises from
the different quantization number in Eq.(\ref{eq:sommer}), from which
we estimate that the energy gap between the lowest two energy levels
at $\gamma=0$ is $\Delta\approx v\sqrt{1-c/v}$, independent of $N$. This agrees 
well with the numerical results in Fig.\ref{fig:gap}.

When $c>v$, the phase space of $H_{\rm mf}$ becomes very different:
there are two local minima with an additional saddle point. In this case,
the Sommerfeld quantization around the two local minima 
gives arise to two sets of Sommerfeld energy levels. In the lower part of 
Fig. (\ref{fig:quant}), for clarity, we have plotted only one set. If two sets 
were plotted, they would form a net of crossings, matching very well with 
the anti-crossing net from $\hh$. In doing the Sommerfeld quantization,
we have ignored the tunneling through the energy barrier between the two 
local minima. Once the tunneling is considered, degeneracies are lifted 
and the crossings become anti-crossings. This shows the energy gaps
inside the triangular net have a different origin from the energy gaps
outside the net or in the case of $c<v$. The energy gaps produced at these 
crossings can be estimated with the WKB method. Since the effective Planck 
constant for $H_{\rm mf}$ is $\hbar/N$, we expect that the gaps decrease
exponentially with $N$. This is exactly what the numerical results in Fig.\ref{fig:gap}
indicates. 

It is now not difficult to understand the tunneling behavior that
we have seen in Fig.\ref{fig:trate}. Let us recall the Landau-Zener 
tunneling in a two-level model\cite{zener}. As $\gamma$ changes with time 
as $\gamma=\alpha t$, the LZ tunneling rate is
$
r_{\rm lz}=\exp\Big(-\frac{\pi \Delta^2}{2\alpha}\Big)\,,
$
where $\Delta$ is the energy gap between the two levels. For a multi-level 
system like our second-quantized model $\hh$, the above equation should 
still be a very good approximation for the tunneling rate between two 
consecutive energy levels. We use the tunneling between the two lowest 
energy levels as an example. As already analyzed, the energy gap changes 
with $N$ as follows, 
\be
\Delta=
\left\{\begin{array}{ll}\displaystyle
\kappa_1 v & c<v\,,\\
N\kappa_2\exp(-\eta N)& c>v\,.
\end{array}\right.
\ee
The parameter $\kappa_1\approx\sqrt{1-c/v}$. The other two parameters
$\kappa_2$ and $\eta$ can be computed with the
WKB method as in Ref.\cite{berry} or with a more sophisticated method\cite{garg}.
This leads to the following tunneling rate 
\be
\displaystyle
r\sim r_{\rm lz}
=\left\{\begin{array}{ll}\displaystyle
\exp\Big(-\frac{\pi \kappa_1^2v^2}{2\alpha}\Big)& c<v\,,\\\displaystyle
\exp\Big(-\frac{\pi N^2\kappa_2^2}{2\alpha e^{2\eta N}}\Big)&c>v\,.
\end{array}\right.
\ee
For the case of $c<v$, it is clear that we have
\be
\lim_{N\rightarrow\infty}\lim_{\alpha\rightarrow 0} r
=\lim_{\alpha\rightarrow 0}\lim_{N\rightarrow\infty}r=0\,,
\ee
which shows that the two limits $\adia$ and $\semi2$ are commutable.
This explains why when $c<v$, both sets of the tunneling rates in 
Fig.\ref{fig:trate} become zero as $\adia$.

For the other case $c>v$, the tunneling rate takes different values 
at two different limits:
\be
\label{eq:limits}
\left\{\begin{array}{l}\displaystyle
\lim_{N\rightarrow\infty}\lim_{\alpha\rightarrow 0} r=0\,,\\
\displaystyle
\lim_{\alpha\rightarrow 0}\lim_{N\rightarrow\infty} r>0\,.
\end{array}\right.
\ee
This reveals that the two limits are no longer commutable.
In the first limit, the adiabatic limit $\adia$ is taken at a fixed number 
of bosons, for which the energy gap is finite and one can always 
be slow enough not to causing tunneling. In the second limit, since
the energy gap is already closed at $\semi2$, tunneling occurs
no matter how slow $\gamma$ changes. This explains why the 
tunneling rate from the mean-field model is always non-zero for $c>v$. 
This incommutability of these two limits also implies
that the mean-field theories, such as Gross-Pitaevskii
equation for BECs, can be invalid for the adiabatic limit. One example
is  the Bloch states
for a BEC in an optical lattice is studied. In such a system,
the Bloch wavenumber $k$ can be regarded as an adiabatic parameter. 
If the Gross-Pitaevskii equation were always valid in the adiabatic limit, 
it would mean that stable Bloch states should exist for all possible $k$.
However, as shown in Ref.\cite{dyn}, a significant portion of
Bloch states are unstable.

\begin{figure}[!htb]
\includegraphics[width=6.5cm]{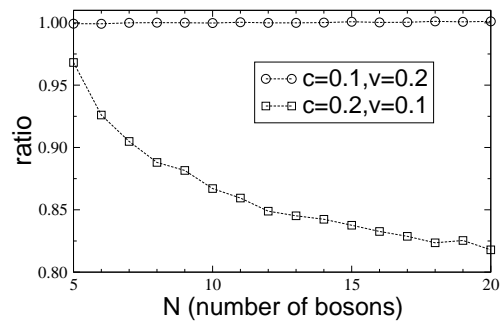}
\caption{Ratio of the bosons in the right well at the
end of tunneling process. The computation is done with the 
second-quantized model $\hh$ with a sweeping rate $\alpha=0.0001$.}
\label{fig:rate}
\end{figure}

With cold atomic gases, we believe that it is now possible to test experimentally
the commutability between the semiclassical limit and the adiabatic limit. 
For instance, one can load a BEC into a double-well potential generated by 
two laser beams\cite{shin}. The energy offset $\gamma$ can be created by using 
different intensity for the two laser beams. To keep the mean-field interaction 
parameter $c$ constant for different boson numbers, one can either use the
Feshbach resonance technique\cite{feshbach} to adjust the interaction between 
atoms or change the size of the trap. In experiments, it is hard
to measure the tunneling rate $r$ between the two lowest energy levels 
as we just discussed. However, one can easily measure the number of atoms
in either of the two potential wells. Once the experiment is set up,
the most striking observation will be as shown in Fig.\ref{fig:rate}.
For $c<v$, if the system initially
has all its atoms in the left well (quantum state $a$), then all the atoms
will be in the right well for a fixed but very small sweep rate. It
does not depend on $N$. For $c>v$, not all the atoms will fall into
the right well: larger $N$ less atoms in the right well. This
difference illustrate our theoretical results on the commutability
of the two limits, $\adia$ and $\semi2$.

J.L. is supported by NNSF of China (10474008). B.W. is supported by 
the ``BaiRen'' program of the Chinese Academy of Sciences and
the 973 project (2005CB724508).

\end{document}